%% file: note468.tex
\documentclass[11pt]{article}
\usepackage{graphicx}

\newcommand{\BABARPubYear}    {02}

\newcommand{\BABARConfNumber} {028}
\newcommand{\SLACPubNumber} {9313}

\input babarsym

\def\eeul      {\ensuremath {3.3\times10^{-7}}}
\def\mmul      {\ensuremath {2.0\times10^{-7}}}
\def\emul      {\ensuremath {2.1\times10^{-7}}}
\def\ldata     {\ensuremath {54.4 \invfb}}
\def\bll {\ensuremath {\Bz \to \ell^{+} \ell^{-}}}
\def\bee {\ensuremath {\Bz \to e^{+} e^{-}}}
\def\bmm {\ensuremath {\Bz \to \mu^{+} \mu^{-}}}
\def\bem {\ensuremath {\Bz \to e^{\pm} \mu^{\mp}}}

\def\mes {\ensuremath {m_{ES}}}
\def\de        {\ensuremath {\Delta E}}
\def\ie {{\it i.e.}}
\def\etal {{\it et al.}}
\def\prd  #1 #2 #3 {Phys.~Rev.~D~{\bf#1},\ #2 (#3)}
\def\plb  #1 #2 #3 {Phys.~Lett.~B~{\bf#1},\ #2 (#3)}
\def\sbaEE{{5.285}}
\def\sbiMM{{5.274}}
\def\sbiEE{{5.273}}
\def\SbiEE{{-0.105}}
\def\SbaEE{{0.050}}
\def\sbaMM{{5.285}}
\def\SbiMM{{-0.050}}
\def\SbaMM{{0.050}}
\def\SbaEM{{0.050}}
\def\sbaEM{{5.284}}
\def\SbiEM{{-0.070}}
\def\sbiEM{{5.274}}
\def\gbiEE{{5.200}}
\def\gbaEE{{5.260}}
\def\GbaEE{{0.400}}
\def\GbiEE{{-0.400}}
\def\gbiMM{{5.200}}
\def\gbaMM{{5.260}}
\def\gbiEM{{5.200}}
\def\GbiMM{{-0.400}}
\def\GbaMM{{0.400}}
\def\gbaEM{{5.260}}
\def\GbaEM{{0.400}}
\def\GbiEM{{-0.400}}
\def\mesresoSPDFSREE{\ensuremath {  3.0\pm  0.2}}

\def\mesresoSPDFSREM{\ensuremath {  2.7\pm  0.1}}

\def\mesresoSPDFSRMM{\ensuremath {  2.6\pm  0.1}}

\def\deresoSPDFSREE{\ensuremath { 29.3\pm  0.9}}

\def\deresoSPDFSREM{\ensuremath { 26.8\pm  0.4}}

\def\deresoSPDFSRMM{\ensuremath { 24.7\pm  0.3}}

\long\def\inst#1{\par\nobreak\kern 4pt\nobreak
    {\it #1}\par\vskip 10pt plus 3pt minus 3pt}

\begin{document}
{\pagestyle{empty}

\begin{flushright}
\babar-CONF-\BABARPubYear/\BABARConfNumber \\
SLAC-PUB-\SLACPubNumber \\
July 2002 \\
\end{flushright}

\par\vskip 5cm

\begin{center}
\Large \bf Search for Decays of {\boldmath $B^0$} Mesons into Pairs of Leptons
\end{center}
\bigskip

\begin{center}
\large The \babar\ Collaboration\\
\mbox{ }\\
July 25, 2002
\end{center}
\bigskip \bigskip

\begin{center}
\large \bf Abstract
\end{center}
We present a  search for the decays $B^0 \rightarrow
e^+ e^-$, $B^0 \rightarrow \mu^+\mu^-$, and $\bem$
in data collected  at the $\FourS$ with the  \babar\ detector at the
SLAC $B$ Factory. Using a data set  of \ldata, we find no evidence for 
a signal  and set the following  preliminary upper limits  at the  
$90\%$ confidence level:  ${\cal B}(\bee)< \eeul$, 
${\cal B}(\bmm) < \mmul$, and ${\cal B}(\bem) < \emul$.

\vfill
\begin{center}
Contributed to the 31$^{st}$ International Conference on High Energy Physics,\\ 
7/24---7/31/2002, Amsterdam, The Netherlands
\end{center}

\vspace{1.0cm}
\begin{center}
{\em Stanford Linear Accelerator Center, Stanford University, 
Stanford, CA 94309} \\ \vspace{0.1cm}\hrule\vspace{0.1cm}
Work supported in part by Department of Energy contract DE-AC03-76SF00515.
\end{center}

\newpage
} 

\input authors_ICHEP2002.tex

\section{Introduction}
\label{sec:Introduction}
\setcounter{footnote}{0}

\begin{figure}[hb]
\begin{center}
\scalebox{0.75}{
\includegraphics[width=15.0cm]{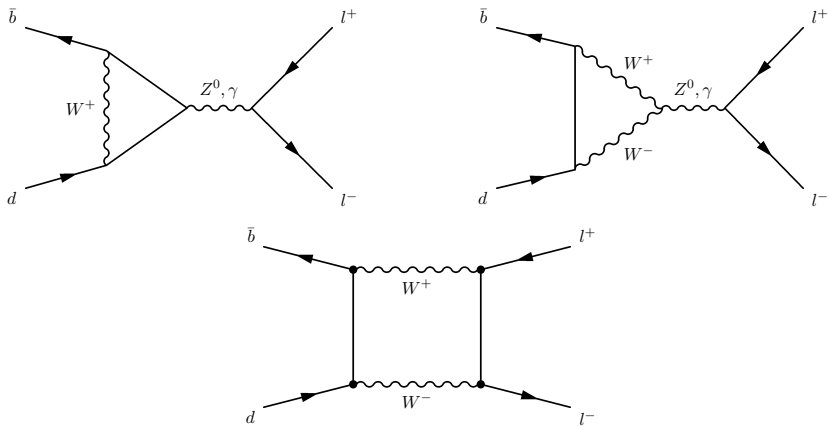}
}
\end{center}
\caption {Feynman diagrams for $\bll$ in the Standard Model. } 
\label{fig:bllfeyn}
\end{figure}

In the Standard Model (SM), rare $B$ decays such as $\bll$,
where $\ell$ refers to $e,\mu$,  are expected to proceed through box and 
loop diagrams as shown in Fig.~\ref{fig:bllfeyn}. 
These decays are highly suppressed since they involve a $b \to d$  
transition and require an internal quark annihilation within the $B$ 
meson which further suppresses the decay by a factor of 
$(f_{B}/M_{B})^2 \approx 2 \times 10^{-3}$ relative to the electroweak
``penguin'' $b \to  d \gamma$ decay. In addition, the decays are
helicity suppressed by factors of $(m_{l}/m_B)^{2}$. $B^0$ decays to leptons 
of two different flavors, \bem, violate lepton flavor conservation and are
therefore strictly forbidden in the SM, although permitted in extensions to 
the SM with non-zero neutrino mass~\cite{numass}. The \bem\ channel includes
both {\ensuremath {\Bz \to e^+ \mu^-}} and {\ensuremath {\Bz \to e^- \mu^+}}.
To date, \bll\ decays have not been observed.
The current best limits from the CLEO~\cite{cleo00} and Belle~\cite{belle01}
collaborations are compared with the SM expectations in 
Table~\ref{tab:expectedBR}. 

\bll\ decays involving tau leptons involve either a soft electron, muon, or 
meson and missing energy (from one or more neutrinos), and require a rather
different search strategy than that presented here for the \bee, \bmm, and
\bem\ channels. The presence of the soft electron or muon in the tau channels 
eliminates them as a source of background to the non-tau channels.

Since these processes are highly
suppressed in the SM, they  are potentially sensitive probes of physics
beyond the SM. Although the branching ratios for these processes are
not significantly enhanced in the Minimal Supersymmetric Standard 
Model (MSSM), various non-minimal supersymmetric models
predict branching  ratios that are significantly larger than those
of the SM. Also,  multi-Higgs-doublet  models  with 
natural  flavor conservation have  extra charged Higgs particles which  
replace the SM $W$-boson in the box diagram of Fig.~\ref{fig:bllfeyn} and 
may enhance  the \bll\ branching  ratios by  an  order of  
magnitude~\cite{NSMTheory}. Similarly, in models with an extra vector-like
down-type  quark~\cite{NirSilverman},  flavor changing neutral currents 
(FCNC)  involving  the $Z^0$  
boson are  induced  due  to  the
different isospin charge of the exotic quark. These models predict the
rate for  $\bll$ to be about  two orders of magnitude  larger than the
expected  SM  rate~\cite{GronauLondon}.  
In addition,  $\bll$  decays are allowed in  specific models  containing
leptoquarks~\cite{PatiSalam}  and supersymmetric  (SUSY) models  without  
$R$-parity conservation.
Furthermore, flavor violating channels such as \bem\
could be enhanced by leptoquarks or
$R$-parity violating operators in SUSY models.

As  shown in Table~\ref{tab:expectedBR},  sensitivity even to models which
produce an  enhancement of a few orders of magnitude to the SM rates for
these  rare decays  is  beyond current  experimental capabilities.  
Observation of a \bll\ decay would, in consequence, provide clear evidence 
for physics beyond the Standard Model.

\begin{table}[hb]
\begin{center}
\caption{The expected branching ratios in the Standard Model~\cite{SMTheory} 
and the current best upper limits (U.L.) in units  of $10^{-7}$  at the
$90\%$  C.L.  from  CLEO~\cite{cleo00}  and Belle~\cite{belle01}.   In
addition,  the measured  number of  events $N_{\rm obs}$  in  the signal
region,  the  expected  background  in  the  signal  region 
$N_{\rm exp}^{\rm bg}$,  and  the
reconstruction efficiency  $\varepsilon$ are quoted.  CLEO's analysis
was  performed on  a data  set corresponding  to  $9.1\invfb$, Belle's
measurement was performed on a data set of $21.3\invfb$.}
\vspace{0.1in}
\begin{small}
\begin{tabular}{|l|l|c|c|c|c|c|c|c|c|} \hline 
Decay & &\multicolumn{4}{c|}{CLEO}&\multicolumn{4}{c|}{Belle}\\\cline{3-6}\cline{7-10}
Mode & SM Expect.       &U.L.&$N_{\rm obs}$&$N_{\rm exp}^{\rm bg}$&$\varepsilon [\%]$ &U.L.  &$N_{\rm obs}$&$N_{\rm exp}^{\rm bg}$ &$\varepsilon [\%]$\\\hline
$e^+e^-$ &$1.9\times10^{-15}$&$8.3$&0&$0.11\pm0.07$&$31.1\pm0.4\pm2.4$&$6.3$&1&$0.6\pm0.8$ &$31.3\pm2.4$  \\ 
$\mu^+\mu^-$ &$8.0\times10^{-11}$&$6.1$&0&$0.22\pm0.07$&$42.4\pm0.5\pm3.2$&$2.8$&0&$0.7\pm0.8$ &$40.0\pm4.3$  \\
$e^{\pm}\mu^{\mp}$ &\multicolumn{1}{|c|}{--}               &$15 $&2&$0.49\pm0.20$&$43.6\pm0.5\pm7.1$&$9.4$&3&$0.7\pm0.8$ &$35.8\pm3.2$  \\ 
\hline
\end{tabular}
\end{small}
\label{tab:expectedBR}
\end{center}
\end{table}

\section{The \babar\ detector and dataset}
\label{sec:babar}

The data used in these analyses were collected with the \babar\ detector
at the \pep2\ $\epem$ storage ring during the years 2000 and 2001.  
The sample corresponds to an integrated luminosity of $54.4~\invfb$ 
accumulated on the \FourS\ resonance (``on-resonance'') and 
$6.4~\invfb$ accumulated at a 
center-of-mass (CM) energy about $40\mev$ below the \FourS\ resonance
(``off-resonance''), which are used for the non-resonant $q\bar q$
background studies.
The on-resonance sample corresponds to $(59.9 \pm 0.7)\times 10^6$ 
\BB\ pairs.  The collider is operated with asymmetric beam energies, 
producing a boost ($\beta\gamma = 0.55$) of the \FourS\ along the 
collision axis.  

\babar\ is a solenoidal detector optimized for the  asymmetric beam
configuration at PEP-II and is described in detail in Ref.~\cite{babarnim}.
The 1.5~T superconducting magnet, whose cylindrical volume is 
1.4~m in radius and 3~m long, contains a charged-particle 
tracking system, a Cherenkov detector (DIRC) dedicated to charged particle 
identification, and a central electromagnetic
calorimeter consisting of 5760 CsI(Tl) crystals.
A forward endcap electromagnetic calorimeter consists of 820 CsI(Tl) crystals.
The segmented flux return, including endcaps, is instrumented with 
resistive plate chambers for muon and \KL identification. This subsystem is
referred to as the instrumented flux return (IFR).
The tracking system consists of a 5-layer 
double-sided
silicon vertex tracker and a 40-layer drift chamber filled with a gas mixture
of helium and isobutane.

\section{Analysis method}
\label{sec:Analysis}

The presence of two charged high-momentum leptons provides 
for a very clean signature for the three decay modes under consideration.
In the CM we require two oppositely-charged 
high-momentum leptons (\ie \ $|p_\ell| \sim m_B/2$) from a common vertex 
consistent with the decay of a $B^0$ meson\footnote{Charge-conjugation is 
implied throughout  this paper.}. 
Since the signal events contain two $B^0$ mesons and no additional particles,
the energy of each $B^0$ in the CM frame must be equal to
the $e^+$ or $e^-$ beam energy. We therefore define
\begin{eqnarray}
m_{\rm ES}&=& \sqrt{(E_{\rm beam}^*)^2 - (\sum_i {\bf p}^*_i)^2} \\
\Delta E&=& \sum_i\sqrt{m_i^2 + ({\bf p}_i^*)^2}- E_{\rm beam}^*,
\end{eqnarray}
where $E_{\rm beam}^*$ is the beam energy in the \FourS\ CM frame,
${\bf p}_i^*$ is the CM momentum of particle $i$ in the
candidate $B^0$-meson system, and $m_i$ is the mass of particle $i$. 
For signal events, the beam-energy-substituted $B^0$ mass, $m_{\rm ES}$, 
peaks at $m_B$. The quantity $\Delta E$ is used to determine
whether a candidate system of particles has total energy consistent with
the beam energy in the CM frame. We require that the beam-energy substituted 
mass, $m_{ES}$, be very close to the mass of the $B^0$ meson and that 
$\Delta E$ be close to zero~\cite{babarnim}. 

To remove background from lepton misidentification, we require tight
electron and muon identification criteria. The electron identification
relies on $E/p$, the ratio of energy deposited in the calorimeter to the
momentum of the particle at the origin, the lateral and azimuthal shower
profiles, and the consistency of DIRC Cherenkov angle with an electron
hypothesis. The muon identification relies on the total number of hits in 
the IFR, the distribution of hits in the different layers, the amount of 
energy released in the calorimeter, and the number of interaction lengths 
which the track has traversed. Suppression of background from 
non-resonant $q\bar q$
production is provided by a series of topological requirements.
In particular, we employ restrictions on the overall magnitude of the event 
thrust and on the magnitude of the cosine of the
thrust angle, $\theta_T$, defined as the angle between the thrust axis of
the particles that form the reconstructed $B^0$ candidate and the thrust
axis of the remaining tracks and neutral clusters in the event.
We also cut on the total multiplicity of both charged tracks and neutral
particles by means of the variable $N_{\rm mult}$ defined as 
\begin{eqnarray}
N_{\rm mult} = N_{\rm trk} + {1\over2}N_\gamma,
\end{eqnarray}
where $N_{\rm trk}$ is the
total number of tracks in the event and $N_\gamma$ is the number of photons
found with an energy $E_\gamma > 80$ MeV. We require $N_{\rm mult} \geq 6.0$.
This variable is especially useful
in the rejection of radiative Bhabha events. We also require that the total
energy in the event be less than 11 GeV. This cut is effective in reducing 
background from two photon events.

\begin{figure}[!htb]
\centering
\includegraphics[width=17.5cm]{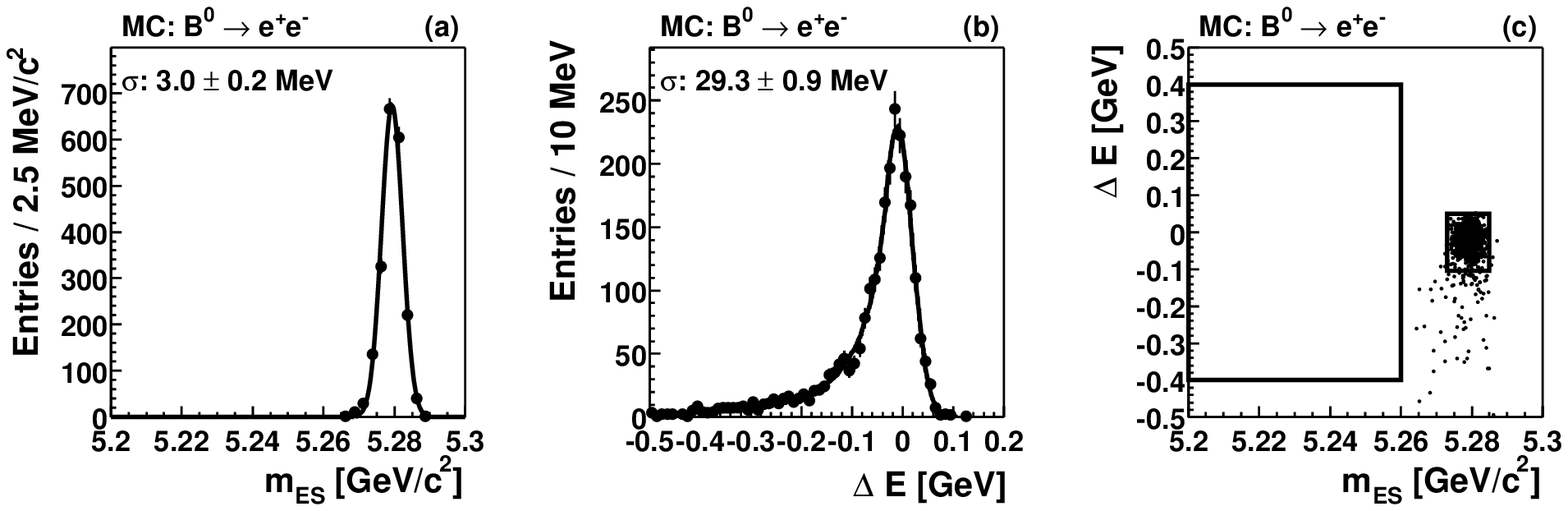}
\includegraphics[width=17.5cm]{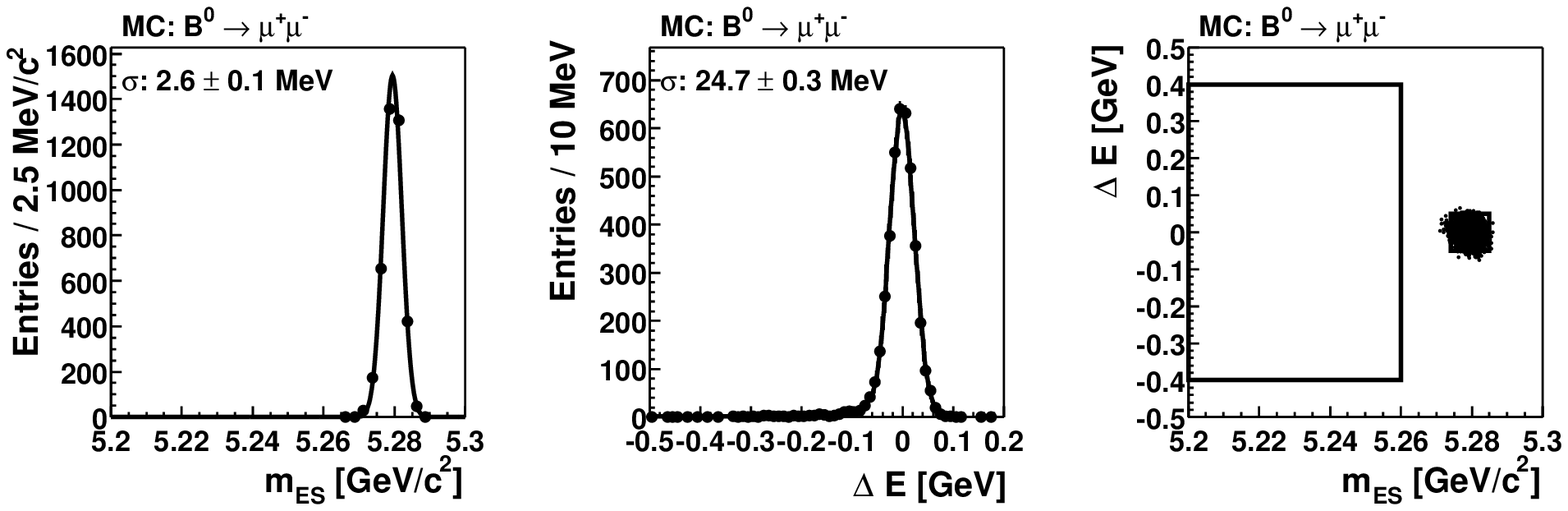}
\includegraphics[width=17.5cm]{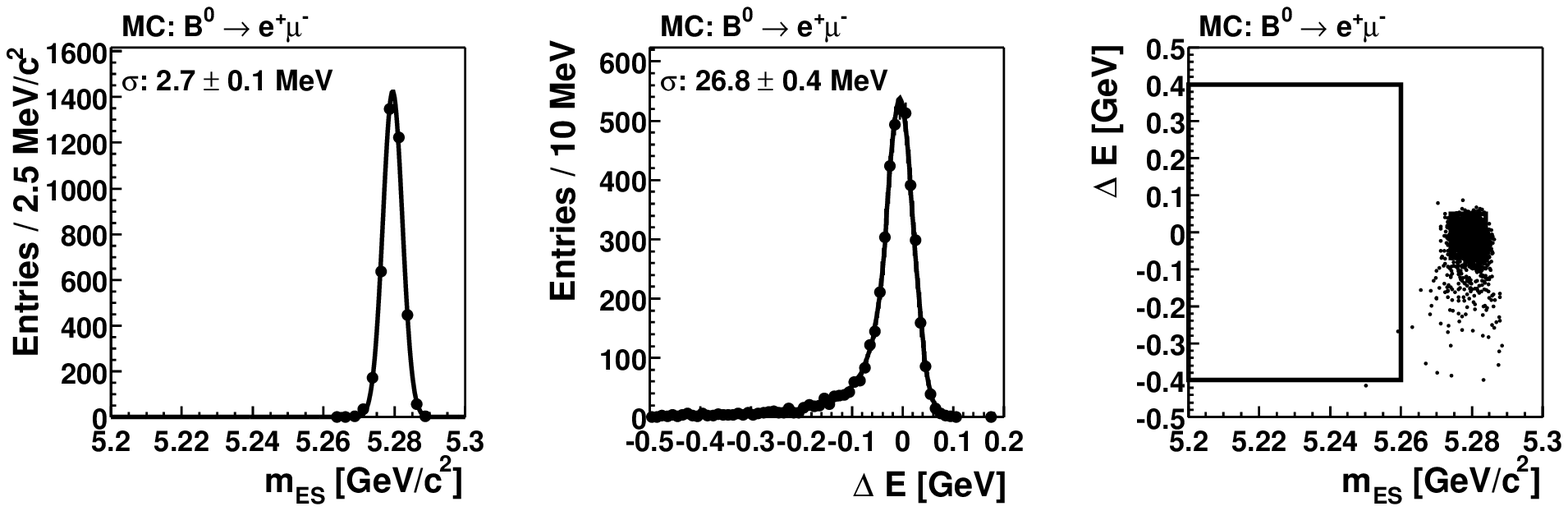}
\caption {Reconstruction of $B$ meson candidates in $\bee$ MC
(top  row), $\bmm$  MC (middle row),  and  $\bem$ MC (bottom  row) with  the
beam-energy  substituted  mass  $m_{ES}$   (a)  and  $\Delta  E$,  the
difference between  the beam  energy and the  energy of  the $B$-meson
candidate in the CM frame (b).
A ``Novosibirsk'' function~\cite{novosibirsk} is used to obtain the widths
of the core  distribution.  Figures (c) show 
the distribution of  $\Delta E$ vs $m_{ES}$.  The smaller box on the right
defines the signal box. The  tail visible  on the
lower  side of  the signal  box is  due to  final state  radiation and
bremsstrahlung. The larger box defines the Grand Sideband (GSB) region.}
\label{fig:breco}
\end{figure}

$\bll$  candidates are selected by simultaneous
requirements on the energy difference $\de$ and the energy-substituted
mass $\mes$. The  size of this ``signal box'' is  chosen to be roughly
$[+2,  -2]\sigma$  of  the  expected  resolution in  $\de$  and  $[+2,
-2]\sigma$ for $\mes$, optimized for the best upper limit.
In  the cases of the $\bee$  and $\bem$ decay modes, the
signal box sizes  in $\de$ are relaxed to  roughly $[+2, -3]\sigma$ and
$[+2,  -2.5]\sigma$, respectively, to accommodate a tail in the distribution 
resulting from final state radiation  and  bremsstrahlung.   
Table~\ref{tab:signaleff} gives the $\mes$ and $\de$ resolutions for the 
three signal channels.
Figure~\ref{fig:breco} illustrates the  $\mes$ and $\de$ distributions
for  Monte Carlo (MC) signal events.   
For the $\de$ distribution, the tail due to final state radiation and 
bremsstrahlung is well described by a ``Novosibirsk'' 
function~\cite{novosibirsk}. Table~\ref{tab:boxtable} summarizes
the $\mes$ and $\de$ cut values used to define the different boxes
in the $(\de, \mes)$ plane.

\begin{table}
\caption{Definition of the signal and Grand Sideband boxes 
in the $(\Delta E, m_{ES})$ plane. The units for $\mes$ are 
[GeV/c$^2$]  and the units for $\de$ are [GeV].}
\begin{center}
\begin{small}
\begin{tabular}{|l|l|l|l|l|l|l|} \hline
        &\multicolumn{2}{c|}{$\bee$}&\multicolumn{2}{c|}{$\bmm$}&\multicolumn{2}{c|}{$\bem$}\\ \hline
Box Name    &\multicolumn{1}{|c|}{$\mes$}  &\multicolumn{1}{|c|}{$\de$}    &
\multicolumn{1}{|c|}{$\mes$}  &\multicolumn{1}{|c|}{$\de$}  &
\multicolumn{1}{|c|}{$\mes$}  &\multicolumn{1}{|c|}{$\de$} \\ \hline
Signal Box                &\sbiEE--\sbaEE  &\SbiEE--\SbaEE &\sbiMM--\sbaMM  &\SbiMM--\SbaMM &\sbiEM--\sbaEM  &\SbiEM--\SbaEM\\
Grand Sideband      &\gbiEE--\gbaEE  &\GbiEE--\GbaEE &\gbiMM--\gbaMM  &\GbiMM--\GbaMM &\gbiEM--\gbaEM  &\GbiEM--\GbaEM\\
\hline
\end{tabular}
\end{small}
\end{center}
\label{tab:boxtable}
\end{table}

\begin{figure}[h]
\centering
\includegraphics[width=5.7cm]{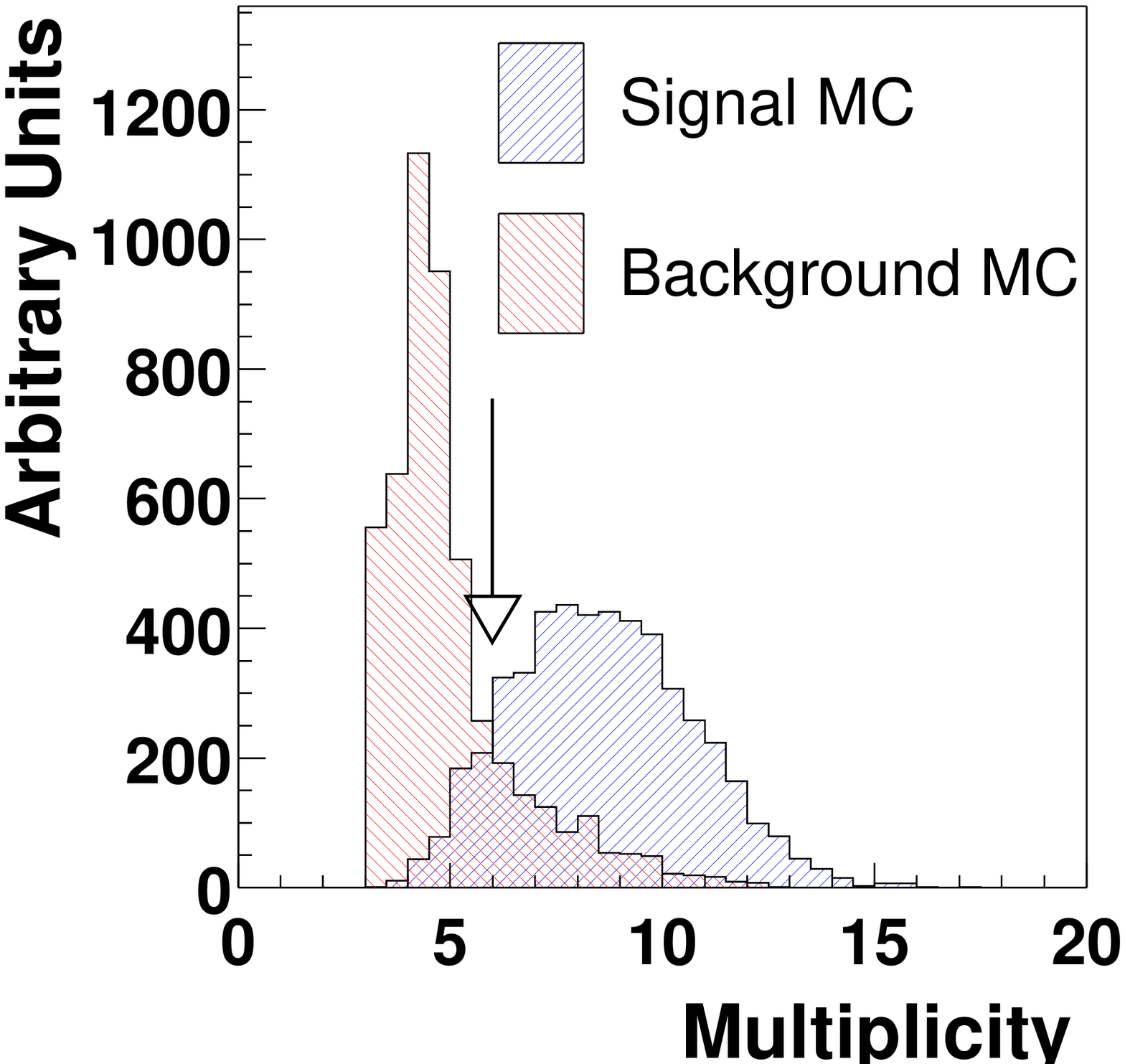}\includegraphics[width=5.7cm]{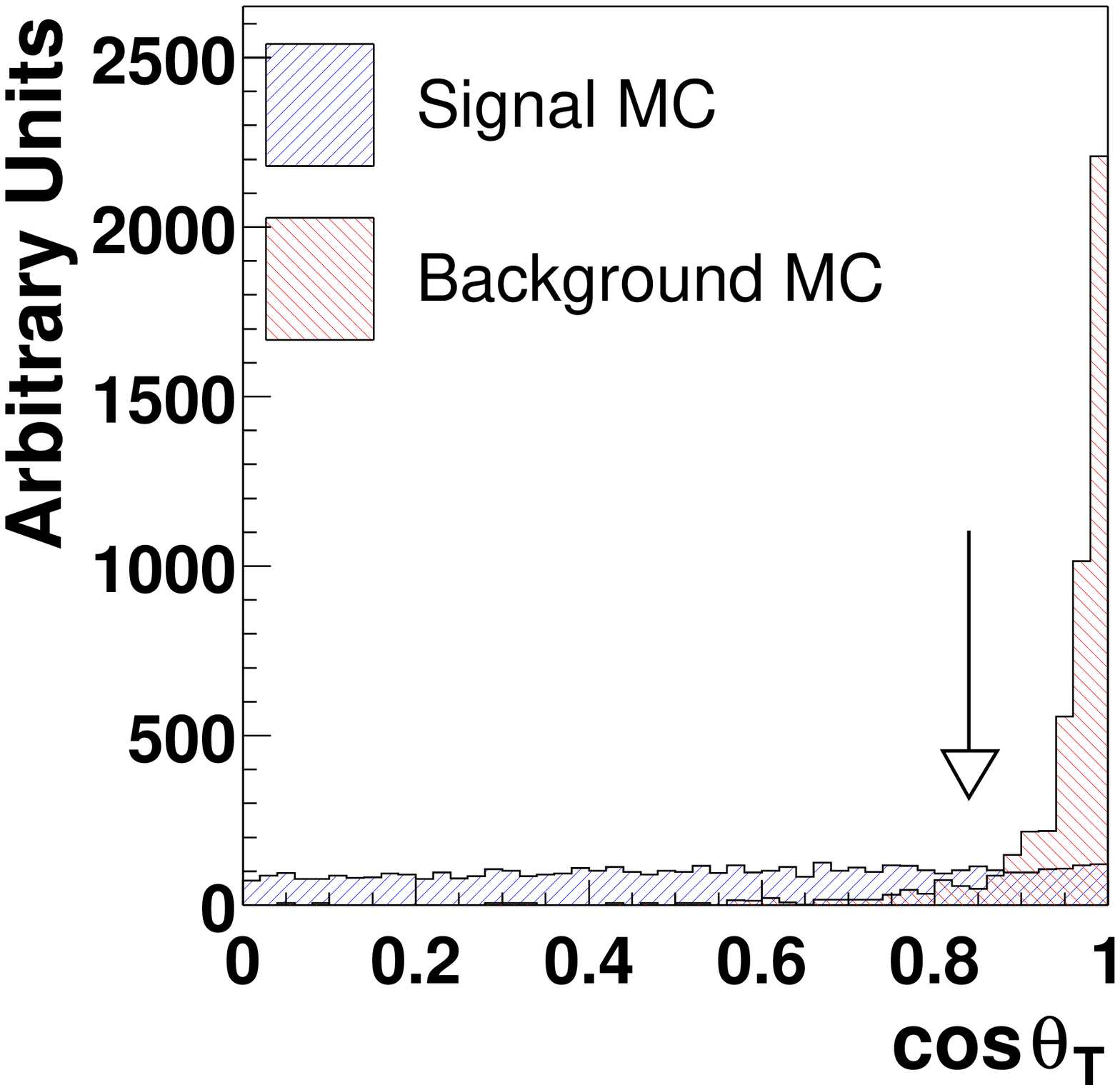}\includegraphics[width=5.7cm]{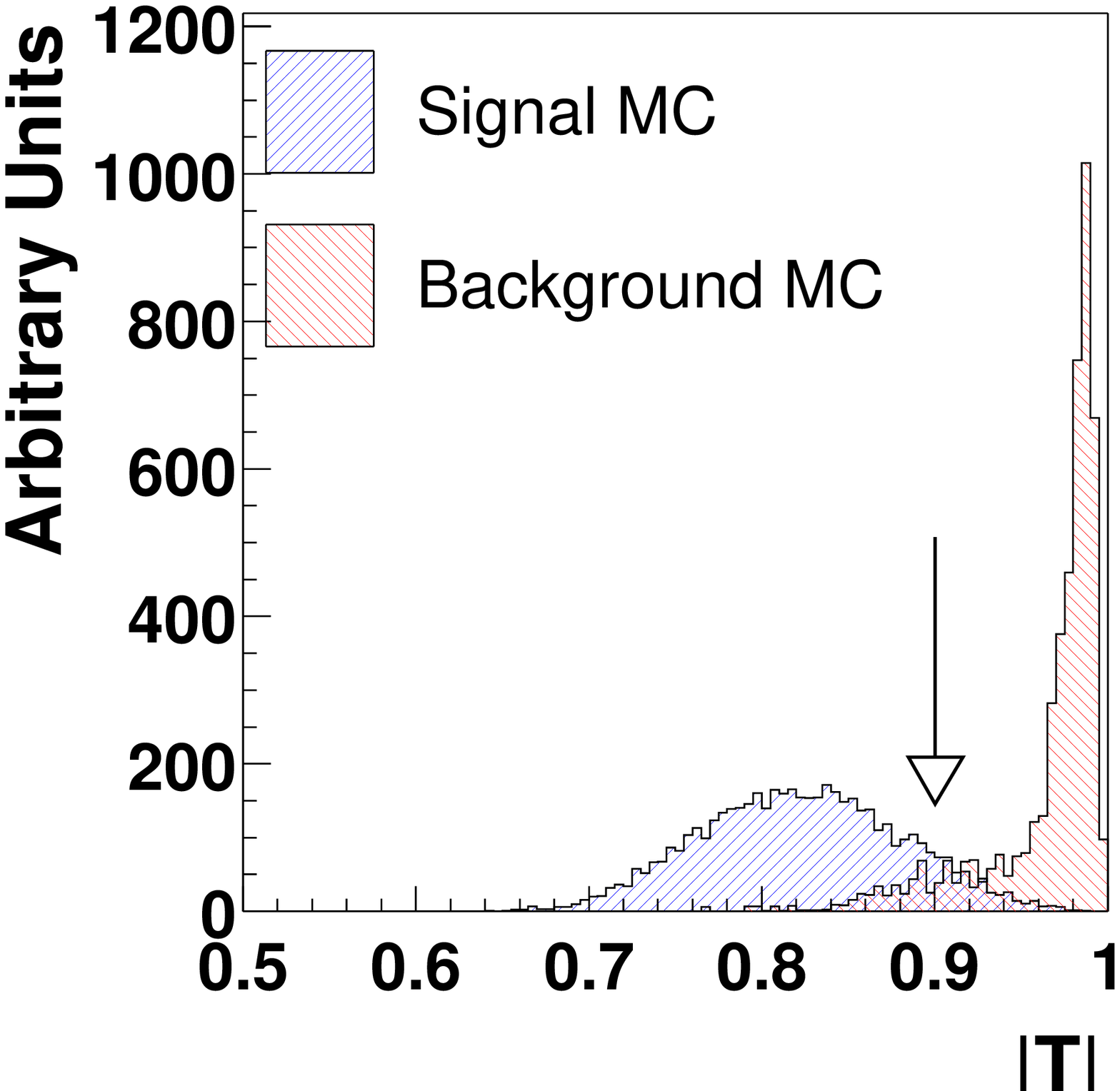}
\caption {Distributions of the main cuts in signal MC and background
MC. The histograms are normalized to equal area. The cut values are
indicated by arrows. The composition of the background MC is explained
in the text.}
\label{fig:cuts}
\end{figure}

We also chose to optimize the cuts on the magnitude of the overall event 
thrust 
$|T|$ and $|\cos \theta_T|$ simultaneously (due to the large correlation). 
The optimal selection criteria for all three channels
was found to be $|\cos \theta_T| < 0.84$ and $|T| < 0.9$.
Figure~\ref{fig:cuts}  illustrates the distributions  of the
multiplicity and event shape variables  in signal and background MC, which
are dominated by non-resonant $\ccbar$- and
$uds$-continuum  processes, but  also include small components from
$\bbbar$ and $\tau$ events. All selection criteria have been 
applied except for the cut on the variable illustrated.  
The efficiencies of the full selection are given in 
Table~\ref{tab:summary}. The  systematic  error  on  the
efficiency is determined by a comparison of the control sample 
$\Bz\to   J/\psi    K_S^0$,    with   $J/\psi\to    e^+e^-$ for $\bee$
and  $J/\psi\to    \mu^+\mu^-$ for $\bmm$, respectively in  data and  MC 
simulation. These comparisions found the dominant uncertainty on
the signal efficiency to be the resolution and scale of \de, 
contributing 4.4\% and 2.6\% for the \bee\ and \bmm\ channels respectively.
Since there is no appropriate control sample for the $\bem$ 
channel, we conservatively set it to be equal to
the largest error obtained from the systematic study for the other channels.
This  yields systematic errors  for all  of the  main cuts  except the
multiplicity cut.  The systematic  error associated with the remaining
cuts is  determined with  a comparison of  signal MC samples  based on
{\tt GEANT3}~\cite{geant3} and {\tt GEANT4}~\cite{geant4}. These two 
simulations employ different material models where the latter is considered
to be more accurate.

\begin{table}[h]
\begin{center}
\caption{$\mes$ and $\de$ resolutions for the three signal channels.}
\vspace{0.1in}
\begin{tabular}{|c|c|c|c|} \hline 
                  & $\bee$                           & $\bmm$                            & $\bem$ \\ \hline
$\sigma(\mes)$ [MeV/c$^2$]& \mesresoSPDFSREE & \mesresoSPDFSRMM  & \mesresoSPDFSREM \\
$\sigma(\de)$  [MeV]& \deresoSPDFSREE  & \deresoSPDFSRMM   & \deresoSPDFSREM  \\ \hline
\end{tabular}
\label{tab:signaleff}
\end{center}
\end{table}

We estimate the background level in the signal box from the data.  The
background expectation is dominated  by different sources in the three
channels. For the $\bee$ channel, the background expectation  is dominated 
by  pairs of true electrons from $\ccbar$ events and two-photon events. 
For the $\bmm$ channel,  about $50\%$ of the total background is due to  
misidentified hadrons  (in combination with  a real  muon).
Two-photon  
processes do not contribute to the background for this channel.
For the $\bem$ channel, the background is composed of real electrons and fake
muons.  Two-photon processes contribute strongly to the background. 

To compare background distributions in data and Monte Carlo event samples,
we use the ``Grand Sideband'' (GSB) box as defined in Table~\ref{tab:boxtable}.
This box is also  used to  estimate  the  functional behavior  
of the $\Delta  E$ dependence   of  the background.

We estimate the background in the signal box assuming that it is
described by the ARGUS function~\cite{argus2} in $\mes$ 
and an exponential function in
$\de$. The shape parameters of the functions are determined in two
different ways: from fitting the data sidebands and from fitting a high
statistics fast Monte Carlo $c\bar c$ sample. 
In both  cases, we determine the normalization in 
the grand sideband and derive the number, $N_{\rm sigBox}$, of expected 
background as follows: 
\begin{eqnarray}
N_{\rm sigBox} &= {\int_{\rm sigBox} f(m_{ES}) \, dm_{ES} \over \int_{\rm GSB} f(m_{ES})\,  dm_{ES}}
\times  {\int_{\rm sigBox}  g(\Delta E)\,d(\Delta E) \over \int_{\rm GSB} g(\Delta E)\,d(\Delta E)}
\times N_{\rm GSB}
\end{eqnarray}
where $N_{\rm GSB}$ is the number of background events found in the
GSB, and $f$ and $g$ are the shapes as determined by the ARGUS and 
exponential fits.
The total background expectations for the three channels are given in 
Table~\ref{tab:summary}.

The actual contents of the signal box were not revealed until the selection
criteria and systematic error estimates were frozen. This technique, often 
referred to as a {\it blind} analysis, is adopted to avoid possible 
experimenter bias.

\section{Results}
\label{sec:Physics}

\begin{figure}
\centering
\includegraphics[width=7.5cm]{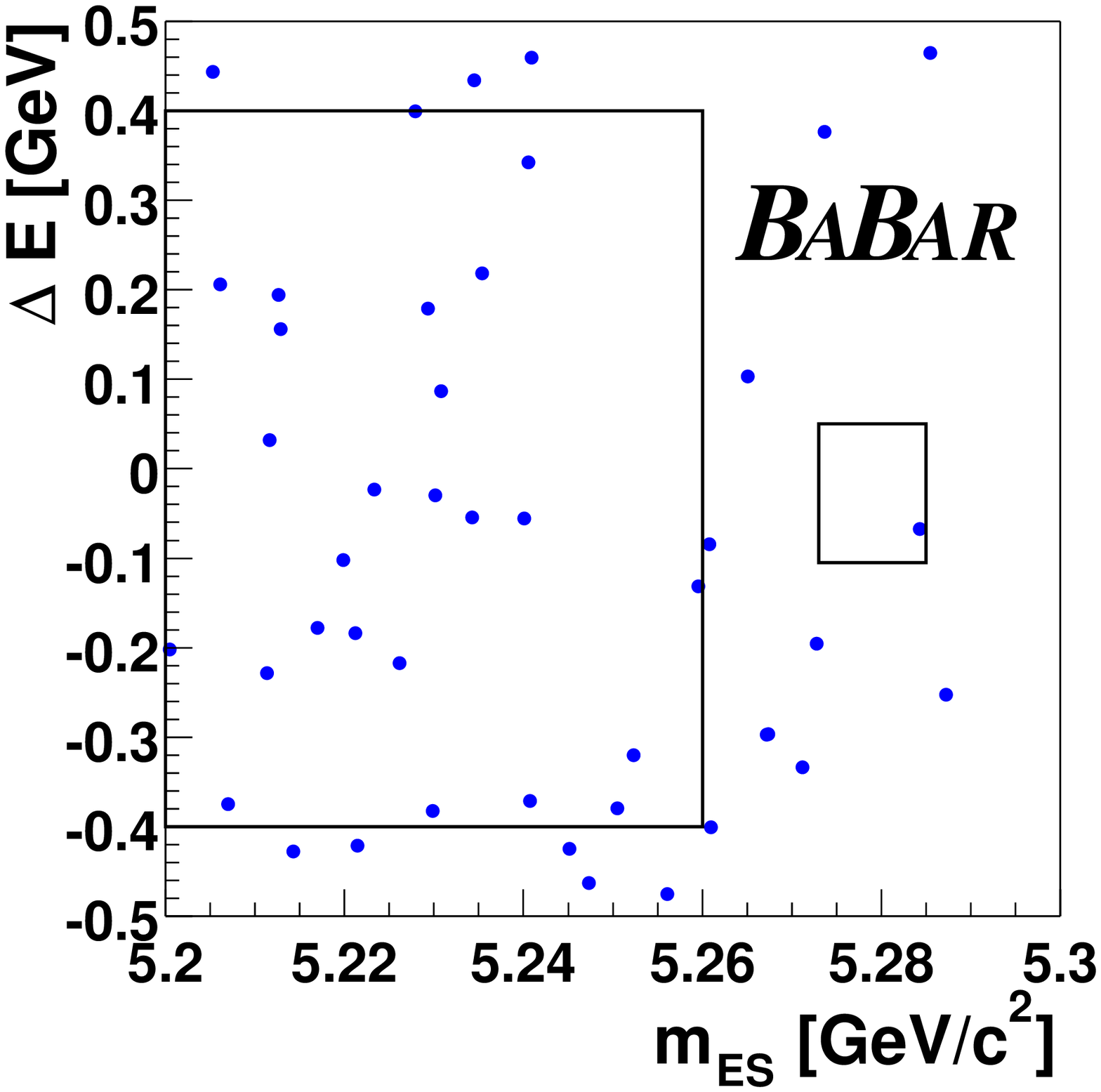}
\includegraphics[width=7.5cm]{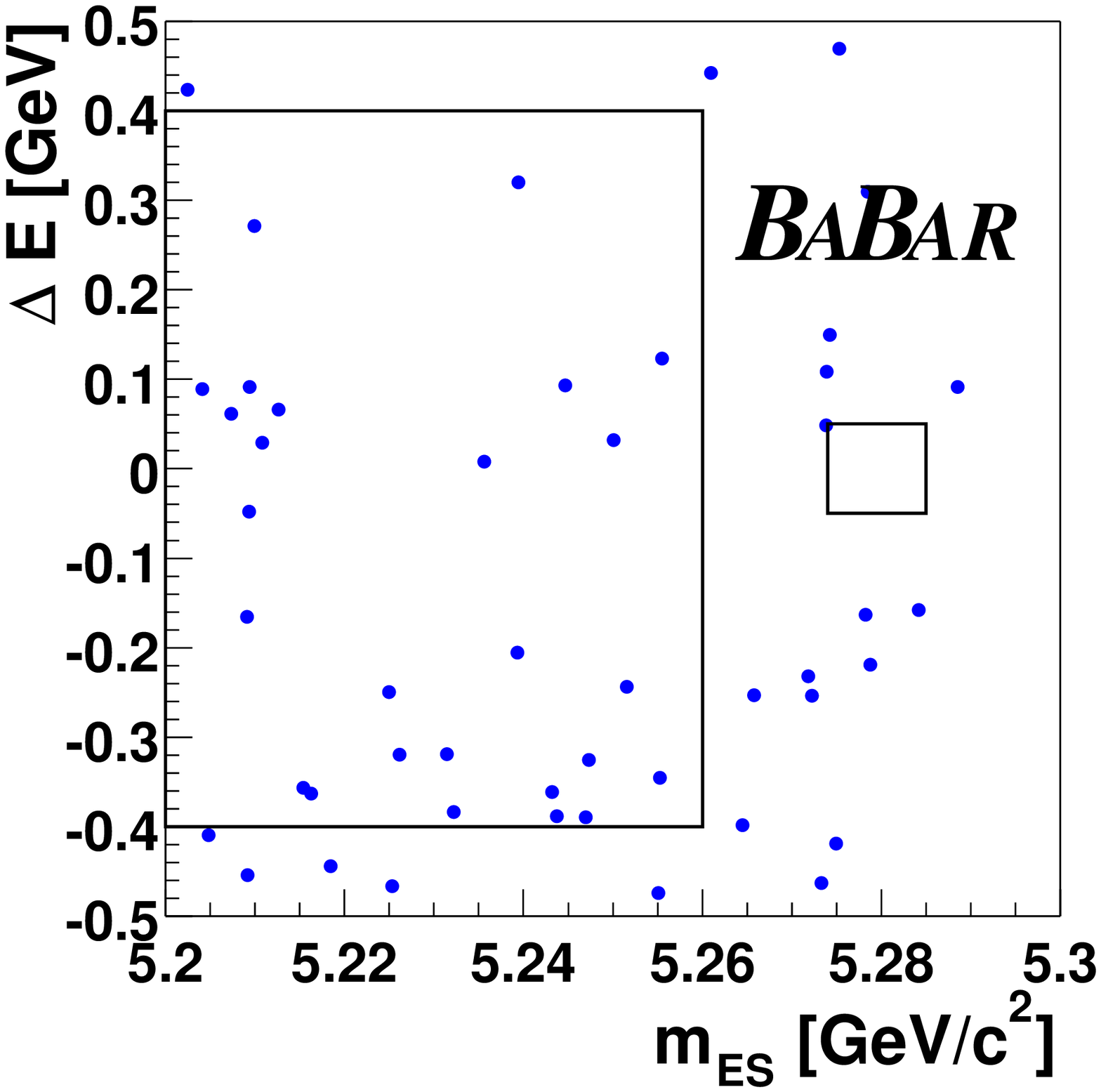}
\includegraphics[width=7.5cm]{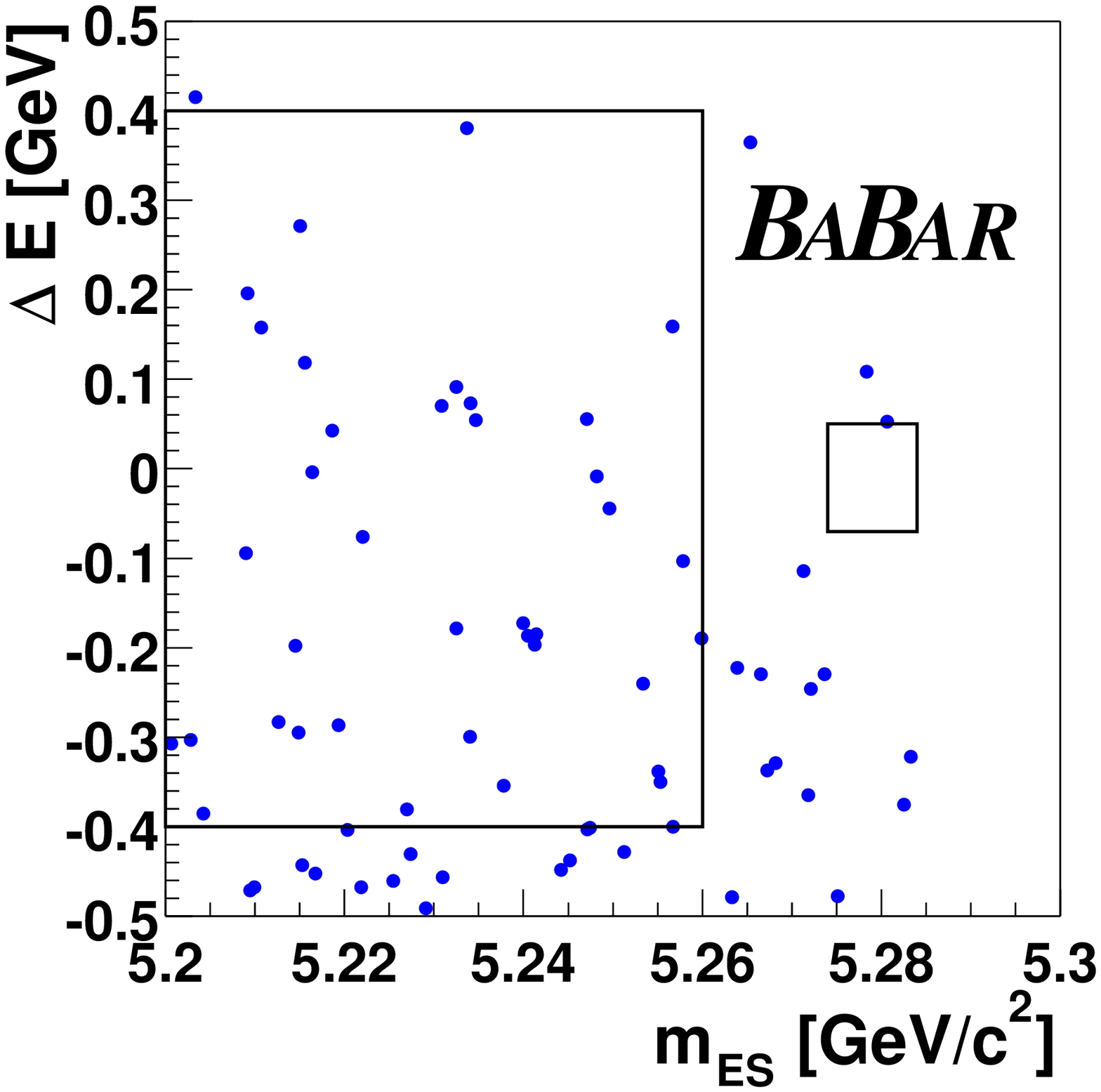}
\caption {Distributions from data of $\mes$ vs $\de$ for $\bee$ (top left), 
$\bmm$ (top right), and $\bem$ (bottom).}
\label{fig:unblind}
\end{figure}

When the contents of the signal box were revealed, one event 
was found in the $\bee$ channel and no events were found in
the other channels as summarized in Table~\ref{tab:summary}.
The $(\mes, \de)$ distributions from data for the three channels
are shown in Fig.~\ref{fig:unblind}.
As can be seen from Table~\ref{tab:summary}, the number of events 
found in the signal box are  compatible with  the  expected background.

We do not perform background  subtraction for the determination of the
upper limit of the branching fraction.  Not yet accounting for the systematic
uncertainties, the upper limit on the branching fraction 
${\cal B}_{\rm UL}({\rm stat})$ is
calculated  as  
${\cal B}_{\rm UL}({\rm stat}) =  N_{UL}/S = N_{UL}/(\varepsilon\times
(N_{\Bz}  + N_{\Bzb}))$,  where $N_{UL}$  is  the upper  limit on  the
number    of   observed    events   $N_{\rm obs}$    and   $S$    is   the
sensitivity. $N_{\Bz} +  N_{\Bzb}$ is equal to the  number of $\FourS$
decays,  since  we assume  equal  production  of  $B^+$ and  $B^0$  in
$\FourS$ decays.

In order to include our systematic uncertainty in the determination of
the upper limit, we follow the prescription  given in 
Ref.~\cite{cousinshighland}.
Assuming a  normal distribution for  the uncertainty  in $1/S$,  the systematic
uncertainty is  accounted for  by convolving the  Poisson probability
distribution for the assumed  branching fraction with a Gaussian error
distribution for  $1/S$. 
The systematic uncertainty on the signal efficiency is found to be 
8.2\% for the $\bee$ channel,
where the main contribution comes from the modeling of the $\mes$ and $\de$
resolutions, and from the uncertainty in the efficiency of the multiplicity 
$N_{\rm mult}$ requirement.
For the $\bmm$ channel, the systematic uncertainty on the signal efficiency 
was found to be $4.7\%$, 
with the primary contribution again being from the modeling of the 
$\mes$ and $\de$ resolutions. 
For the $\bem$ channel, the systematic uncertainty 
on the signal efficiency was taken to be the same as for the
$\bee$ channel.
The total uncertainty was calculated by summing in quadrature the 
systematic uncertainty
on the number of \BB events, the signal efficiency, and the statistical error 
on the signal efficiency.

As summarized in Table~\ref{tab:summary}, the resulting preliminary upper 
limits for ${\cal B}(\bee)$, ${\cal B}(\bmm)$, and ${\cal B}(\bem)$ are 
$3.3\times10^{-7}$, $2.0\times10^{-7}$, and $2.1\times10^{-7}$ respectively.

\begin{table}
\begin{center}
\caption{Summary  of the analyses.  $N_{\rm exp}$  is the  number of  expected
signal events,  assuming a branching fraction  of $1.9\times10^{-15}$ for the 
\bee\ channel and $8.0\times10^{-11}$ for the \bmm\ channel. $N_{\rm obs}$
is the number  of observed events in the signal  box.  $N_{\rm exp}^{\rm bg}$
is the expected number of background events in the signal box. 
${\cal B}_{\rm UL}(B^0\to \ell^+\ell^-)$ is the  upper limit on the branching 
ratio at the $90\%$ C.L.}
\vspace{0.1in}
\begin{small}
\begin{tabular}{|c|c|c|c|c|c|} \hline
channel&$N_{\rm exp}$ &$N_{\rm obs}$   &$N_{\rm exp}^{\rm bg}$ &$\varepsilon [\%]$ &${\cal B}_{\rm UL}(B^0\to \ell^+\ell^-)$ \\ \hline

$\bee$&$1.9\times10^{-8}$ &$1$ &$0.60\pm0.24$ &$19.3\pm0.4_{\rm stat}\pm1.6_{\rm sys}$ &$3.3\times10^{-7}$\\

$\bmm$&$3.2\times10^{-2}$ &$0$ &$0.49\pm0.19$ &$18.8\pm0.3_{\rm stat}\pm2.0_{\rm sys}$&$2.0\times10^{-7}$ \\

$\bem$&\multicolumn{1}{|c|}{--} &$0$ &$0.51\pm0.17$ &$18.3\pm0.4_{\rm stat}\pm1.5_{\rm sys}$ &$2.1\times10^{-7}$\\

\hline
\end{tabular}
\end{small}
\label{tab:summary}
\end{center}
\end{table}

\section{Acknowledgments}
\label{sec:Acknowledgments}


\input acknowledgements


\end{document}

%% file: authors_ICHEP2002.tex
\begin{center}
\small

The \babar\ Collaboration,
\bigskip

B.~Aubert,
D.~Boutigny,
J.-M.~Gaillard,
A.~Hicheur,
Y.~Karyotakis,
J.~P.~Lees,
P.~Robbe,
V.~Tisserand,
A.~Zghiche
\inst{Laboratoire de Physique des Particules, F-74941 Annecy-le-Vieux, France }
A.~Palano,
A.~Pompili
\inst{Universit\`a di Bari, Dipartimento di Fisica and INFN, I-70126 Bari, Italy }
% ...

\end{center}\newpage

%% file: acknowledgements.tex
We are grateful for the 
extraordinary contributions of our \pep2\ colleagues in
achieving the excellent luminosity and machine conditions
that have made this work possible.
The success of this project also relies critically on the 
expertise and dedication of the computing organizations that 
support \babar.
The collaborating institutions wish to thank 
SLAC for its support and the kind hospitality extended to them. 
This work is supported by the
US Department of Energy
and National Science Foundation, the
Natural Sciences and Engineering Research Council (Canada),
Institute of High Energy Physics (China), the
Commissariat \`a l'Energie Atomique and
Institut National de Physique Nucl\'eaire et de Physique des Particules
(France), the
Bundesministerium f\"ur Bildung und Forschung and
Deutsche Forschungsgemeinschaft
(Germany), the
Istituto Nazionale di Fisica Nucleare (Italy),
the Research Council of Norway, the
Ministry of Science and Technology of the Russian Federation, and the
Particle Physics and Astronomy Research Council (United Kingdom). 
Individuals have received support from 
the A. P. Sloan Foundation, 
the Research Corporation,
and the Alexander von Humboldt Foundation.

%% file: note468.bbl
\begin{thebibliography}{99}
\bibitem{novosibirsk} \babar\ defines the Novosibirsk function to be \\
$f(E) = A\cdot \exp\left[-{1\over2}
\left({\log(1+\tau(E-\nu)\cdot{\sinh(\tau\sqrt{\log4})\over \sigma\tau\sqrt{\log4}})\over\tau}\right)^2+\tau^2\right],$ \\ 
where $\tau$ is the ``tail parameter'' 
(describing how  much is contained in the tail), $\sigma$ is the width, and 
$\nu$ is the peak position.
\end{thebibliography}
